\documentclass[10pt,letterpaper]{article}
\usepackage[top=0.85in,left=1in,footskip=0.75in,marginparwidth=2in]{geometry}

\usepackage[utf8]{inputenc}

\usepackage{nameref,hyperref}

\usepackage[right]{lineno}

\usepackage{microtype}
\DisableLigatures[f]{encoding = *, family = * }

\raggedright
\usepackage{changepage}
\usepackage{cite}
\usepackage[aboveskip=1pt,labelfont=bf,labelsep=period,singlelinecheck=off]{caption}

\makeatletter
\renewcommand{\@biblabel}[1]{\quad#1.}
\makeatother

\usepackage{lastpage,fancyhdr,graphicx}
\usepackage{epstopdf}
\usepackage{color}
\usepackage{amsmath}
\definecolor{Gray}{gray}{.25}

\usepackage{graphicx}
\usepackage{amsfonts}
\usepackage{sidecap}
\usepackage{longtable}
\usepackage{booktabs}
\usepackage{wrapfig}
\usepackage[pscoord]{eso-pic}
\usepackage[fulladjust]{marginnote}
\reversemarginpar

\begin{document}
\vspace*{0.35in}

\begin{flushleft}
{\Large
\textbf\newline{Graph-GIC: A Smart and Parallelized Geomagnetically Induced Current Modelling Algorithm Based on Graph Theory for Space Weather Applications}
}
\newline
\\
Wen Chen\textsuperscript{1,2},
Ding Yuan\textsuperscript{1,2,*},
Xueshang Feng\textsuperscript{1,2},
Stefaan Poedts\textsuperscript{3,6},
Zhengyang Zou\textsuperscript{4},
Song Feng\textsuperscript{5},
Yuxuan Zhu\textsuperscript{1,2},
Tong Yin\textsuperscript{1,2}
\\
\bigskip
\bf{1} Shenzhen Key Laboratory of Numerical Prediction for Space Storm, School of Aerospace, Harbin Institute of Technology, Shenzhen, Guangdong 518055, China
\\
\bf{2} Key Laboratory of Solar Activity and Space Weather, National Space Science Center, Chinese Academy of Sciences, Beijing, China
\\
\bf{3} Centre for Mathematical Plasma Astrophysics, Department of Mathematics, KU Leuven, B-3001 Leuven, Belgium
\\
\bf{4} State Key Laboratory of Lunar and Planetary Sciences, Macau University of Science and Technology, Macau, China
\\
\bf{5} Faculty of Information Engineering and Automation, Kunming University of Science and Technology, Kunming 650500, China
\\
\bf{6} Institute of Physics, University of Maria Curie-Sklodowska, ul.\ Marii Curie-Sklodowskiej 1, 20-031 Lublin, Poland
\\
\bigskip
* yuanding@hit.edu.cn

\end{flushleft}
\begin{abstract}
Geomagnetically Induced Current (GIC) refers to the electromagnetic response of the Earth and its conductive modern infrastructures to space weather and would pose a significant threat to high-voltage power grids designed for the alternative current operation. To assess the impact of space weather on the power grid, one needs to calculate the GIC on a national or continental scale. In this study, we developed a smart and parallelized GIC modelling algorithm, \emph{Graph GIC}. This algorithm deploys a graph representing a power grid in a single-line diagram, in which substations/transformers act as nodes and transmission lines as edges. With these denotations, a power grid and its electric parameters are mathematically represented with an adjacency matrix and an admittance matrix. We used sparse matrix and parallelisation techniques to expedite the intensive computation in cases of large-scale power grids. The \emph{Graph GIC} was validated with a benchmark grid, applied to the GIC calculation of the 500 kV power grid of Guangdong, China, and conducted preliminary analysis on the grid's susceptibility to geomagnetic storms. The \emph{Graph GIC} algorithm has the advantage of an intuitive and highly scalable graph representation of a power grid at any scale. It achieves high-accuracy calculation and a speedup of about 18 times after parallelisation. This algorithm could be applied to assess the impact of space weather on a power grid up to continental scales and could be incorporated into  global space weather modelling frameworks.
\end{abstract}


\section{Introduction}
\label{sec:intro}

During a space weather storm, coronal mass ejection (CME) is usually launched into the heliosphere, carrying a huge volume of ionized particles and electromagnetic fields. These materials would disrupt the Earth's magnetic field and magnetosphere-ionosphere, known as a geomagnetic storm. A geomagnetic storm would induce electric currents in the Earth and modern technological systems, such as pipelines, high-speed railways, and power systems \cite{pirjola2000,liu2016,Kappernman1990,pilipenko2021}. Since geomagnetically induced current (GIC) fluctuates over a typical timescale of minutes to hours \cite{lundby1985}, they could be considered quasi-steady direct currents (DC). Under the interruption of GIC, a power transformer is shifted to a nonlinear operation regime and suffers the half-cycle saturation effect, which would deposit significant heat on the metal core. This process would lead to a rise in temperature in the core and cooling oil, and the temperature rise caused by GIC may fail a transformer \cite{molinski2002,price2002,marti2012}. GIC tends to accumulate at substations, particularly in coastal areas, regions with significant variations in geological conductivity, and geomagnetic anomaly zones, where the increase in GIC is more likely \cite{pulkkinen2012,boteler1998,ngwira2013}. In March 1989, a key power transformer was burned in the power grid of Quebec, Canada, and this event led to a nine-hour power outage in Canada \cite{boteler2019}.

To calculate the GIC in a power grid, one has to simplify the AC power grid into a single-line-diagram \cite{pirjola2000, horton2012}. The commonly used methods for calculating GIC in power grids are the Network Impedance Matrix method and the Nodal Admittance method (NAM). Traditionally, GIC simulation regards substations as holistic nodes within the power grid and disregard the GIC in transformer windings, focusing only on calculating the GIC at a single voltage level. However, the GIC in the transformer windings has a direct and adverse impact on transformers \cite{albertson1981,kappenman1989}. Lehtinen et al. \cite{lehtinen1985} and Pirjola developed the LP method for the multi-voltage power grid,  which has to add virtual node to denote ungrounded transformer and associate an infinite resistance to it, this step generates numerical difficulties in the matrix calculations. Pirjola et al.\cite{pirjola2022} modified the LP method (LPm) by replacing the network impedance matrix with an earthing admittance matrix to facilitate GIC calculations. This modified method and sparse matrix techniques enable efficient calculations for power networks with large-scale nodes.

In this paper, we propose a more efficient GIC computation method based on the LPm method designed to facilitate the scalability of the power grid. The algorithm leverages sparse matrix and parallel computing to enhance the computational efficiency of GIC time series analysis. We validate the algorithm using the Horton network, demonstrating significant improvements in computational efficiency while maintaining accuracy. Additionally, we calculate and analyze the GIC levels in the Guangdong power grid during two magnetic storm events.
\section{Modeling GIC}
\label{sec:gic}

The first step of calculating GIC is to develop a DC equivalent model for the AC power grid. The DC equivalent model assemblies (1) the AC model for power flow analysis, (2) resistance data, such as substations/transformers and transmission lines, and (3) geographic coordinates of the substations/transformers. The second step uses a graph to represent the DC grid model. A substation/transformer is denoted with a node, whereas a transmission line is represented with an edge. The DC equivalent power grid model in a graph could be stored as an adjacency matrix ($\mathbf{A}$). 

The NAM and LP methods are mathematically equivalent and are used at the convenience of the algorithm \cite{boteler2014}, and the LPm method is more collegiate than the LP method \cite{pirjola2022}. This study implemented the LPm method to calculate the GIC of the power grid's substations/transformers and transmission lines. A graph of a power grid is handled with the NetworkX \cite{NetworkX-python} package based on Python. The connection of nodes (edges) is represented with an adjacency matrix; a resistance matrix was used to store the resistance parameters of the power grid. 

The adjacency matrix $\mathbf{A}$ is a symmetric matrix, with a dimension of $\mathbb{R}^{N\times N}$, where $N$ is the number of nodes. If node $\mathrm{n}$ and $\mathrm{k}$ are connected, then $a_{\mathrm{n}\mathrm{k}}=1$; or otherwise, $a_{\mathrm{n}\mathrm{k}}=0$. For diagonal elements, if the node k is grounded, then $a_{\mathrm{k}\mathrm{k}}=1$.

The resistance matrix $\mathbf{R}$ is stored similarly to the adjacency matrix.  $r_{\mathrm{n}\mathrm{k}}$ is the resistance of the transmission line between node $\mathrm{n}$ and $\mathrm{k}$. If $r_{\mathrm{n}\mathrm{k}}=\infty$, it means node $\mathrm{n}$ and $\mathrm{k}$ are not connected at all, or there is a device in between that blocks the flow of GIC. If $r_{\mathrm{n}\mathrm{k}}= 0$, it means node $\mathrm{n}$ and $\mathrm{k}$ are physically connected by highly conducting media. $r_{\mathrm{k}\mathrm{k}}$ denotes the grounding resistance. If node $\mathrm{k}$ is not grounded, then $r_{\mathrm{k}\mathrm{k}}=\infty$. The admittance matrix $Y$ is defined as the reciprocal of the resistance matrix in a graph, as reactance is negligible in GIC calculation.

\subsection{Lehtinen-Pirjola Method}

Lehtinen et al. \cite{lehtinen1985} proposed the LP method for GIC modelling in multi-voltage power grids and could be used to calculate the GIC of substations and transformers. The LP method starts with Kirchhoff’s law for the currents at a nodal point:
\begin{equation}
    i_k=\sum_{n=1,n\neq k}^N i_{nk}=-\sum_{n=1, n\neq k}^N i_{kn}.\label{eq:i_k_lp}
\end{equation}
The LP method relates the current in a transmission line to the electromotive force, 
\begin{equation}
    i_{kn}=(e_{kn}+v_k-v_n)y_{kn}. \label{eq:i_nk_lp}
\end{equation}
The LP method defines a “perfect earthing” current caused by the driving emf:
\begin{equation}
    J_k^{e}=- \sum_{n=1, n\neq k}^N e_{kn}y_{kn}, \label{eq:je_k} 
\end{equation}
Substituting equations (\ref{eq:je_k}) and (\ref{eq:i_nk_lp}) into equation (\ref{eq:i_k_lp}) becomes
\begin{align}
  i_k& = J_k^{e}- \sum_{n=1,, n\neq k}^N (v_k-v_n)y_{kn} \label{eq:i_k_je}   \\
     & =J_k^{e} - v_k\sum_{n=1, n\neq k}^N y_{kn} + \sum_{n=1, n\neq k}^N v_n y_{kn}.
\end{align} 
If we define a new admittance matrix $\mathbf{Y}^n$, with the diagonal elements as 
\begin{equation}
    Y^n_{kk} = \sum_{n=1, n\neq k}^N y_{nk},
\end{equation}
and the off-diagonal elements as
\begin{equation}
    Y^n_{kn} =-  y_{kn}.
\end{equation}
\eqref{eq:i_k_je} could be written in matrix format,
\begin{equation}
    \mathbf{I}^{e} = \mathbf{J}^{e} - \mathbf{Y}^{n} \mathbf{V}^n, \label{eq:i_vector}
\end{equation}
where $\mathbf{I}^{e}$, $\mathbf{J}^{e}$, and $\mathbf{V}^n$ are the column vectors of currents, perfect earthing current and nodal voltages at all nodes. Here $\mathbf{V}^n$ is related to the earthing impedance matrix $\mathbf{Z}^{e}$.
\begin{equation}
  \mathbf{V}^n = \mathbf{Z}^{e} \mathbf{I}^{e}.\label{eq:v_vector}
\end{equation}
In this definition, the LP method is related to the nodal voltage differently from the NAM method,
\begin{equation}
   v_k=\sum_{n=1}^N Z_{kn}^e i_n.
\end{equation}
By inserting equation (\ref{eq:v_vector}) into equation (\ref{eq:i_vector}), we obtain the current from node to ground as the only unknown variable
\begin{equation}
    \mathbf{I}^{e} = (\mathbf{1}+\mathbf{Y}^{n}\mathbf{Z}^{e})^{-1}\mathbf{J}^{e},  \label{eq:i_vector_lp}
\end{equation}
where $\mathbf{1}$ is the unit matrix. 

If we relate the nodal voltage with the source current directed to the earth as done in the NAM method, by inserting equation (\ref{eq:v_vector}) into equation (\ref{eq:i_vector_lp}), and resulting 
\begin{equation}
   \mathbf{V}^{n}  = [(\mathbf{Z}^{e})^{-1}+\mathbf{Y}^{n}]^{-1}\mathbf{J}^{e}.  \label{eq:V_vector_lp}
\end{equation}

\eqref{eq:V_vector_lp} is the same used in NAM method only if $\mathbf{Z}^{e}$ is a diagonal matrix, meaning that the voltage at node $k$ only generates a current flow from node $k$ to the earth, but not via other grounding nodes. This interaction between nodes is only true when they are very close, as demonstrated in \cite[]{Pirjola2008}. If the earthing current at any node does not affect the voltages at the other nodes, $\mathbf{Z}^{e}$ becomes a diagonal matrix related to the earthing resistance. Then we define a ground admittance matrix $\mathbf{Y}^{e}$ as the inverse of $\mathbf{Z}^{e}$, then equation (\ref{eq:V_vector_lp}) becomes
\begin{align}
    \mathbf{V}^{n} & = (\mathbf{Y}^{e}+\mathbf{Y}^{n})^{-1}\mathbf{J}^{e} \label{eq:v_YJ}\\ 
    & = \mathbf{Y}^{-1}\mathbf{J}^{e}.  
\end{align}

Here, $\mathbf{Y}$ is defined precisely as done in NAM method. Thus, the LP method is mathematically equivalent to the NAM method. 

\subsection{Modified Lehtinen-Prijola Method}

Since the LP method requires consideration of an ungrounded substation/transformer by creating a virtual node, Pirjola et al. \cite{pirjola2022} modified Lehtinen-Prijola method (LPm method), as shown in equation (\ref{eq:v_YJ}). The matrix $\mathbf{Y^e}$ represents the earthing admittance matrix of the nodes, characterized by non-zero elements only on the diagonal. 
\begin{equation}
\mathbf{Y}_{nk}^{e}=\begin{cases}
	     y_k=\frac{1}{r_k}  & n=k   \\
             0 & n \neq k
		   \end{cases}
    \label{eq:Y_ije}
\end{equation}

This method involves the inversion of the network admittance matrix $(\mathbf{Y}^{n})$ and the earthing admittance matrix $(\mathbf{Y}^{e})$ using the LPm method. Since $[\mathbf{Y}^{n}+\mathbf{Y}^{e}]$ is a positive-definite symmetric matrix, it can be efficiently decomposed using the Cholesky decomposition, as in equation (\ref{eq:cholesky}),
 \begin{equation}
     \mathbf{L}\mathbf{L^T}=(\mathbf{Y}^{n}+\mathbf{Y}^{e}) \label{eq:cholesky}
 \end{equation}
Combine the equations (\ref{eq:cholesky}) and (\ref{eq:v_YJ}), and obtain,
\begin{equation}
    \mathbf{L}\mathbf{L^T}\mathbf{V^n} = \mathbf{J^e}  \label{eq:LLV}
\end{equation}
The column matrix $\mathbf{P}$ is defined as equation (\ref{eq:Lp_J}),
\begin{equation}
    \mathbf{L}\mathbf{P}=\mathbf{J^e}  \label{eq:Lp_J}
\end{equation}
Then inserted into equation (\ref{eq:Lp_V}) to solve the matrix $\mathbf{V^n}$.
\begin{equation}
    \mathbf{L^T}\mathbf{P}=\mathbf{V^n} \label{eq:Lp_V}
\end{equation}

The LPm method with Cholesky decomposition also provides an efficient method for calculating GIC at multiple time steps. In larger power grid of national or continental scale,  node number reach thousands or even more, Over 99\% of the elements in the admittance matrix is zero. Cholesky factorization could use sparse matrix methods and, hence, could reduce memory use and computation time \cite{STOTT198719,press2007numerical} 

\section{Novel Techniques in the Graph-GIC}

\label{Sec:Optimization}

\subsection{Graph representation for a Power Grid}
The continuous development and optimization of power systems, along with the increasing number of substations and power plants, present significant challenges for power system modelling. To achieve scalability of the power grid, we represent substations and buses in the grid as nodes and transmission lines and transformer windings as edges. The network is mapped to facilitate directionless connections between nodes, allowing for greater flexibility in modelling the system's topology. 

\begin{figure}[ht]
  \centering

    \begin{minipage}[b]{0.9\textwidth}
    \centering
    \begin{picture}(0,0)
      \put(-200,0){\textbf{(a)}}
    \end{picture}
        \includegraphics[width=0.7\textwidth]{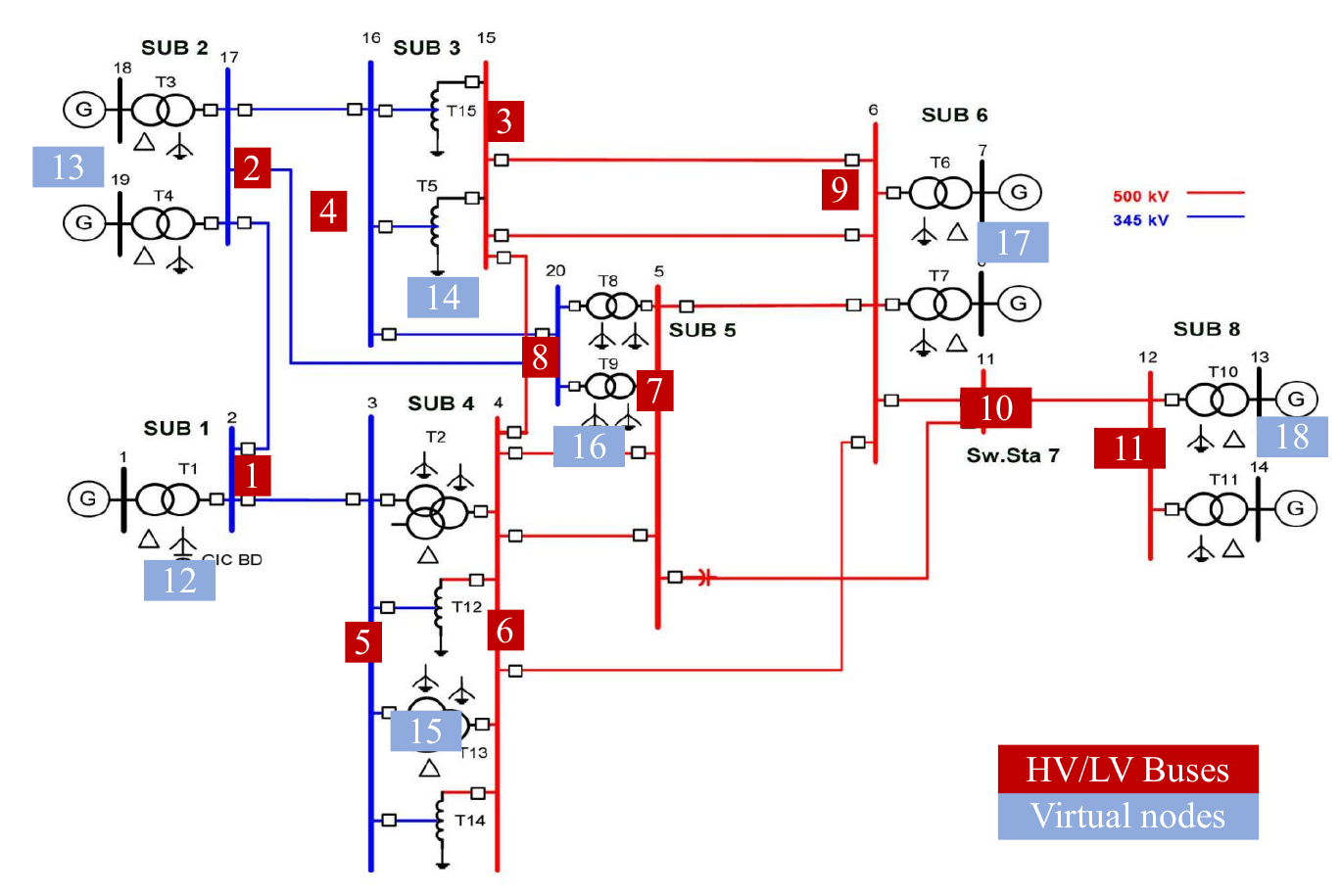}
    \end{minipage}  
\begin{minipage}[b]{0.9\textwidth}
        \centering
        \begin{picture}(0,0)
      \put(-200,0){\textbf{(b)}}
    \end{picture}
        \includegraphics[width=0.7\textwidth]{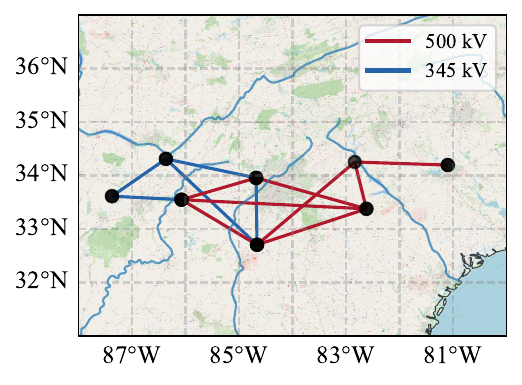}
\end{minipage}
\caption{Single-line diagram of Horton grid used for generating DC equivalent model, HV/LV buses are visualized in red, whereas the neutral nodes are marked with blue. \cite{horton2012} used with permission. (b) A DC equivalent model with substations as nodes and internal connections between the transformers was not displayed. Black nodes signify simplified models of substations, while blue and red lines represent transmission lines of 345 kV and 500 kV voltage levels, respectively.}
    \label{fig:horton}
\end{figure}

Figure \ref{fig:horton}{a} shows the single-line diagram of the Horton grid \cite{horton2012}, which consists of 7 substations, one switch station, and 15 transformers, along with 20 busbars and 15 transmission lines.  The grid is simplified based on the graphical network, containing 18 nodes and 38 line connections. Figure \ref{fig:horton}{b} draws a graph of the power grid overlaid on a geographic map to characterize the substations and transmission lines. In this representation, a substation could accommodate multiple transformers, e.g., SUB 4 consists of four transformers. A graph reppulkkinen2012resentation enables the power grid to be highly scalable by adding any number of nodes (substations/transformers) and edges (transmission lines). Similarly, when expanding a power system, new nodes can be individually modelled and integrated into the existing model without the need to remodel the entire power system.

\subsection{Sparse Matrix Techniques}

Sparse matrix theory is applied to matrices that contain many zero elements. In a graph (matrix) representation of a power grid with thousands of nodes, a node is connected to a few neighbouring nodes, so sparse matrix techniques can be deployed to store power grid parameters and reduce computation time in matrix manoeuvring. 

\begin{figure}[ht]
  \centering
  \begin{minipage}[b]{0.45\textwidth}
    \centering
    \begin{picture}(0,0)
      \put(-100,10){\textbf{(a)}}
    \end{picture}
    \includegraphics[width=\textwidth]{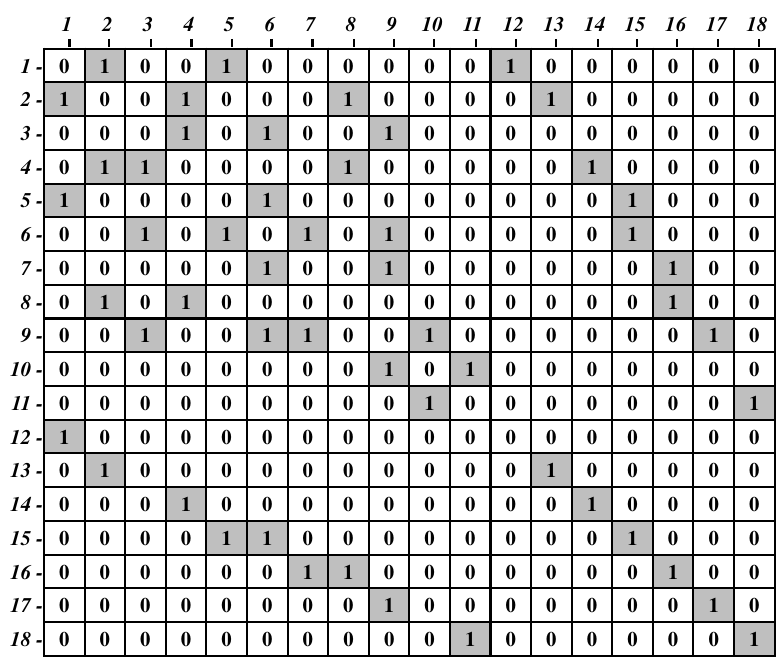}
  \end{minipage}
  \hfill
  \begin{minipage}[b]{0.45\textwidth}
    \centering
    \begin{picture}(0,0)
      \put(-100,0){\textbf{(b)}}
    \end{picture}
    \includegraphics[width=\textwidth]{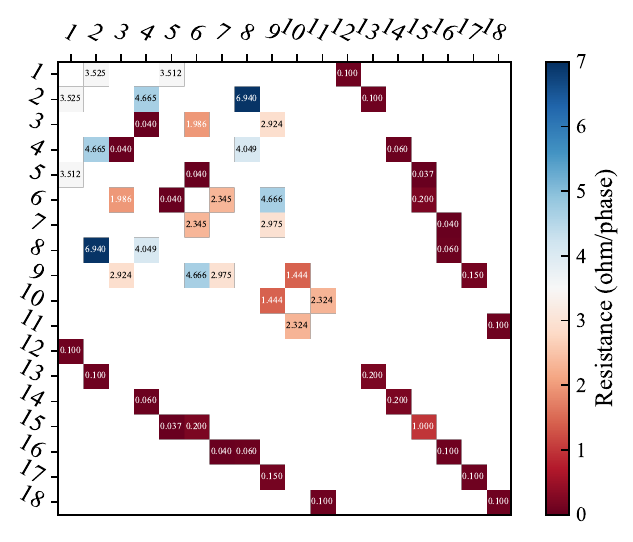}
  \end{minipage}
\caption{(a) Horton benchmark grid in the form of an adjacent matrix. (b) Node network resistance matrix.}
    \label{fig:Matrix}
\end{figure}

In equation \ref{eq:cholesky}, the lower triangular matrix $\mathbf{L}$ can be represented as a sparse matrix and substituted into equation \ref{eq:LLV} to calculate the node voltages efficiently. Sparse matrices have an advantage in large-scale power grids; see Pirjola et al. \cite{pirjola2022}. Figure~\ref{fig:Matrix}{a} visualizes the adjacency matrix ($\mathbf{A}$) of the Horton grid, which intuitively shows the interconnections between nodes. Figure \ref{fig:Matrix}{b} shows the associated resistance between nodes (off-diagonal elements) and earthing resistance (diagonal elements). If there is no connection between node k and node n, or if there is a device that blocks the GIC, then $\mathrm{r_{kn}}$ is effectively infinite and is not represented in the heat map. Similarly, for node k that is not grounded, $\mathrm{r_{kk}}$ is also effectively infinite and is not depicted here either. The node admittance matrix $(\mathbf{Y}^{n})$ and earthing admittance matrix $(\mathbf{Y}^{e})$ could be derived and are consistent with the calculation in Pirjola et al. \cite{pirjola2022}.

\subsection{Paralleling Computing}

To enhance the computational efficiency of GIC under time series, this paper incorporates parallel computing technology, considering the independence of the GIC calculation process for each grid point. This paper compares two parallel computing processes under the LPm method. The first parallel computing process (called LPm1 Process) includes the computation of the matrix $\mathbf{Y^{e}}+\mathbf{Y^{n}}$ as well as the time series of node voltages and GIC. In the second method (called LPm2 Process), the matrix $\mathbf{Y^{e}}+\mathbf{Y^{n}}$ is first computed and used as an input parameter, and the parallel computing process only includes the time series of node voltages and GIC. To evaluate the performance of parallel computing, the maximum number of cores is set to 40 for both, and the speedup ratio of the model was tested across various numbers of processing units. 

The input data consists of geomagnetic fields with a period of 24 hours and a resolution of 1 minute, resulting in a total of 1440 data points. 
\begin{figure}[ht]
    \centering
    \includegraphics[width=0.5\textwidth]{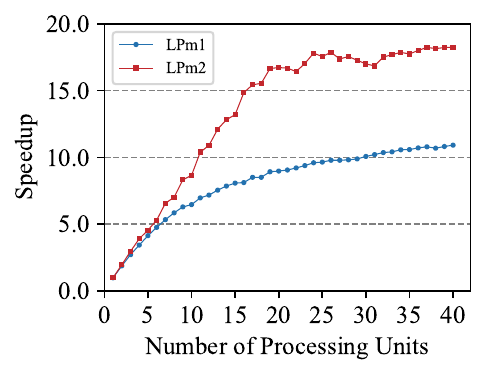}
    \caption{The speedup ratios of the ``Graph-GIC" model for simulating the Horton Grid GIC levels under parallel computations.}
    \label{fig:speedup}
\end{figure}

Figure~\ref{fig:speedup} presents the speedup for two LPm approaches. As the number of processes gradually increases, the speedup for both LPm1 and LPm2 processes slowly increases, with the LPm1 Process reaching a speedup of 11 and  LPm2 reaching a speedup of about 18. This indicates that parallel computing can significantly enhance computational efficiency. Comparing the speedup ratios of the LPm1 Process and the LPm2 Process, it is evident that reducing the serial portion of parallel computation can effectively improve performance.

\section{Algorithm Testing and Validation}
\subsection{Test with Uniform Electric Fields}

To assess the model's feasibility, we assume uniform electric fields of 1 V/km in the northward and eastward directions over the power grid's area.

\begin{figure}[ht]
\centering
\includegraphics[width=0.6\textwidth]{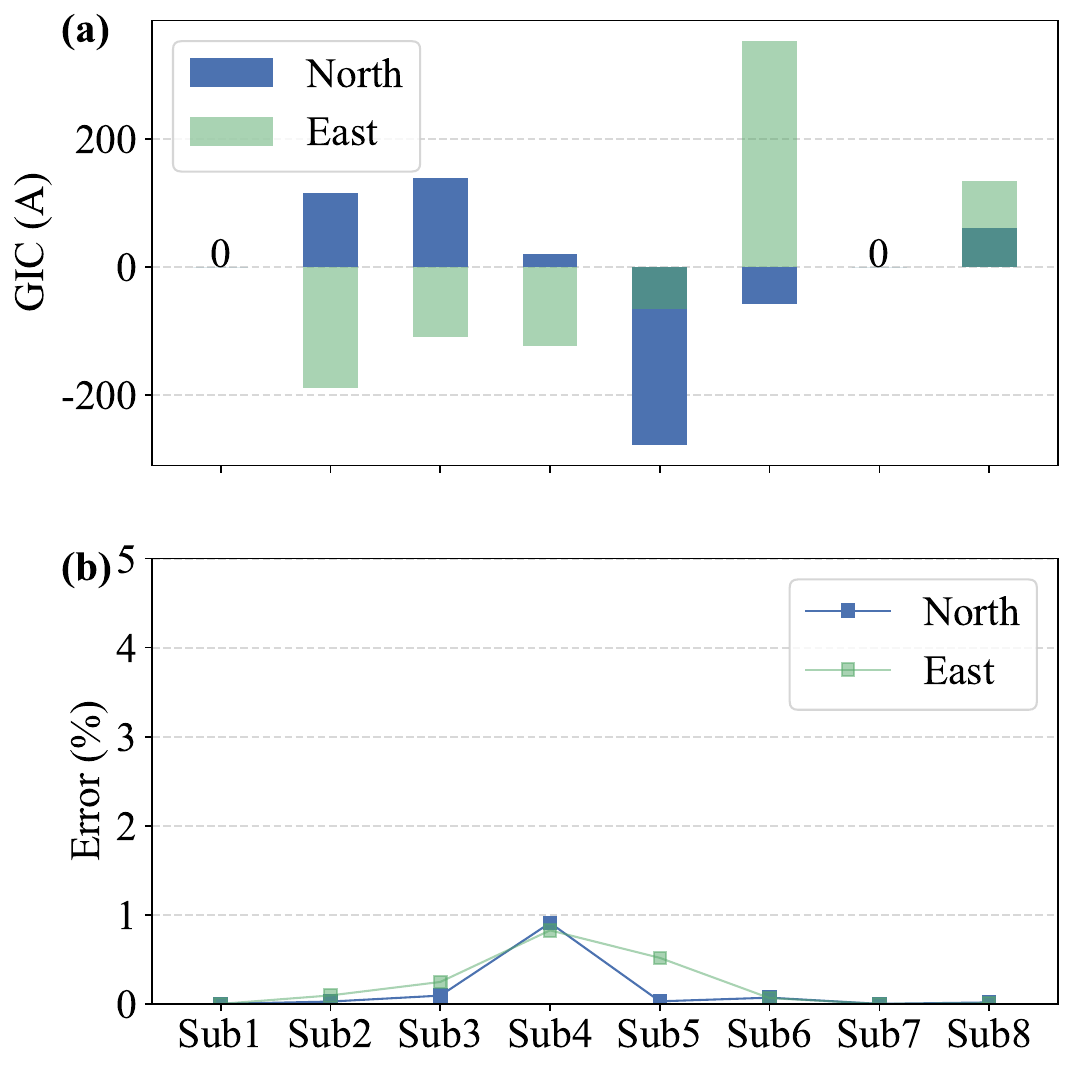}
\caption{(a) geomagnetically unduced currents (GIC) per substation  from ``Graph-GIC" algorithm (three phase). (b) The GIC errors per substation between modelled and Horton data. \label{fig:GIC substations}}
\end{figure}

Figure \ref{fig:GIC substations}{} shows the northward and eastward GIC at each substation (a) and the error (b) concerning the results of Horton et al. \cite{horton2012}. Positive GIC denotes the current flowing from the substation to the ground; negative GIC refers to the current flowing from the ground to the substation. We could see that the \emph{Graph GIC} algorithm reproduced the data with very high accuracy; the maximum error was less than 1\%. We note that substation 1 (SUB 1) and the Sw.Sta 7 have zero GIC, consistent with Horton et al. \cite{horton2012}. 

\begin{figure}[ht]
    \centering
    \includegraphics[width=0.6\textwidth]{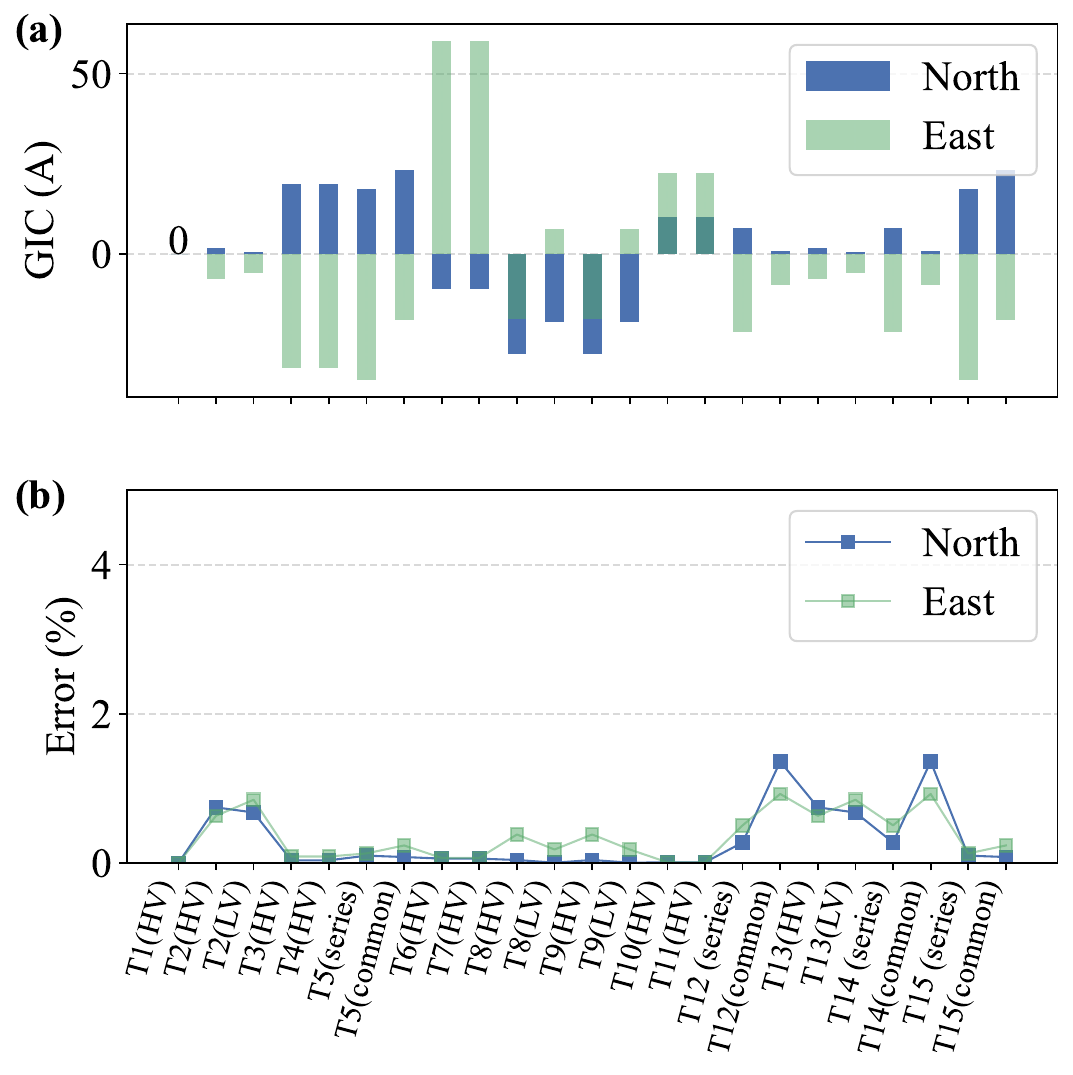}
    \caption{(a) The geomagnetically induced currents (GIC) per transformer from \emph{Graph GIC} algorithm (per phase). (b) The GIC errors per transformer between modelled and Horton data.}
    \label{fig:GIC transformers}
\end{figure}

Figure~\ref{fig:GIC transformers}{} shows the northward and eastward GIC at transformers and the associated error concerning the results of Horton et al. \cite{horton2012}. We could see that most transformers have a GIC with an error of less than 1\%. Only at two transformers the relative error of GIC reached 1.36\%. 

The comparison in both substation and transformer level indicates that the \emph{Graph GIC} algorithm calculated the GIC values in high consistency with Horton et al. \cite{horton2012} model. We, therefore, use the \emph{Graph GIC} algorithm for GIC modelling of the realistic power grid. 

\subsection{Test with Artificial Geomagnetic Storm}

In addition to the uniform electric fields, we modelled the GIC of the Horton et al. \cite{horton2012} model under the impact of an artificial geomagnetic storm. We synthesized an time-varying geoelectric fields with disturbances in both the northward and eastward directions, this step mimic a geomagnetic storm and its sub-storms. The geoelectric field modelled the impact of three geomagnetic pulses and was mathematically given as
\begin{equation}
	E(t) = \begin{cases}
	      0, &  0 \leq t < 2  \\
	      e^{-(t-2)}, &  2 \leq t < 5 \\
       -0.6\times e^{-(t-5)},  &5 \leq t < 7 \\
       0.6\times e^{-(t-7)},   & 7 \leq t < 12.
		   \end{cases}
    \label{eq:electric}
\end{equation}

\begin{figure}[ht]
    \centering
    \includegraphics[width=17cm]{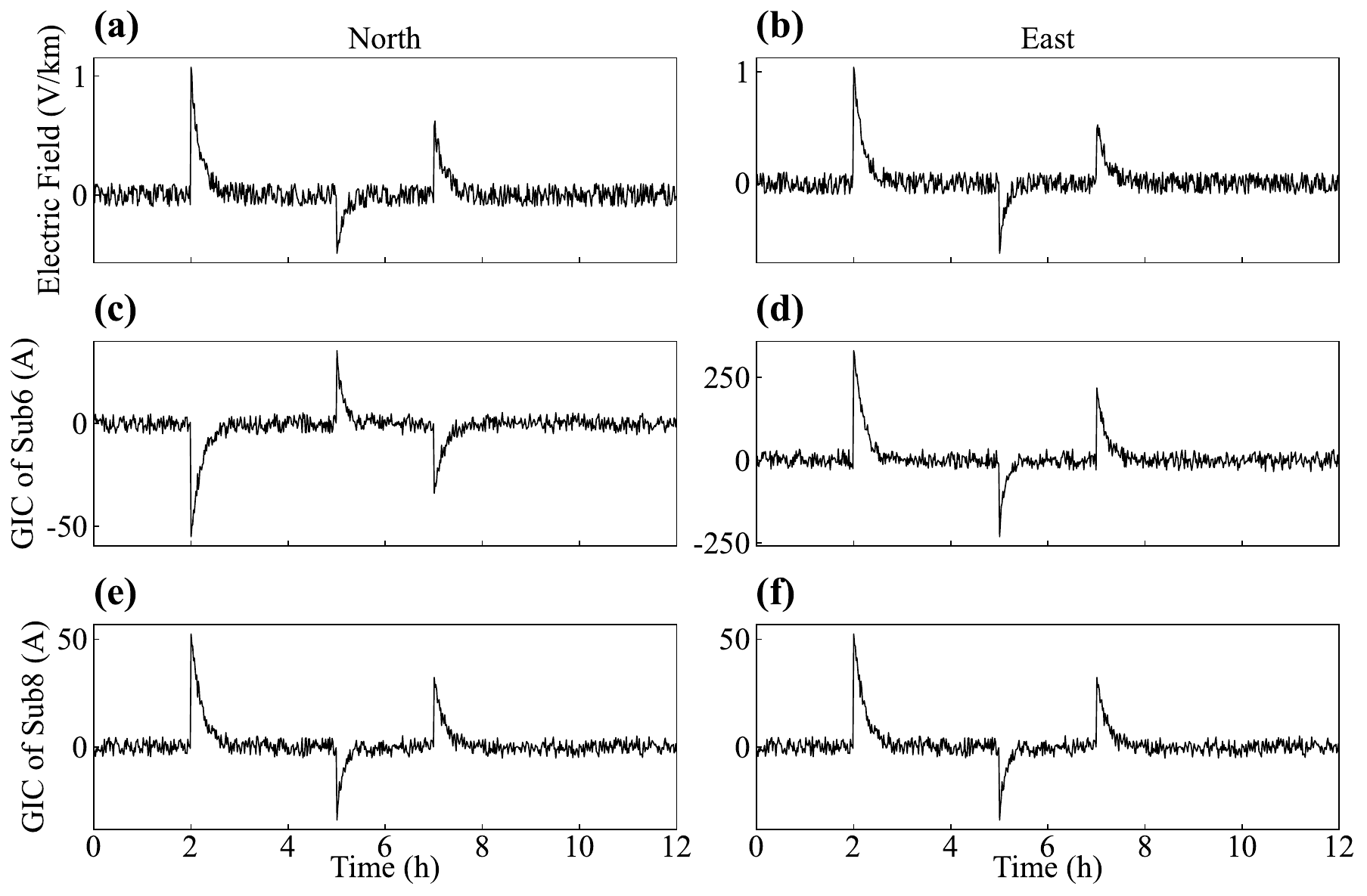}
    \caption{Synthetic geomagnetic storms in forms of northward (a) and eastward (b) geoelectric field. The northward (c,e) and eastward (d,f) GIC in substations 6 and 8 (SUB 6, SUB 8), respectively. }
    \label{fig:GIC_time}
\end{figure}

A Gaussian noise a standard deviation of 0.1 V/km is added to the geoelectric field in order to sythesize the random fluctuations on the earth surface. Figure \ref{fig:GIC_time}{a} and \ref{fig:GIC_time}{b} present northwards (X) and eastwards (Y) geoelectric fields as input to the Horton et al. \cite{horton2012} power grid model. This geoelectric field synthesises the impact of three consecutive geomagnetic storms with a duration of about 30 minutes to 1 hour.

Figures~\ref{fig:GIC_time}{c} and \ref{fig:GIC_time}{e} plot the northward GIC at substation 6 (SUB 6) and substation 8 (SUB 8), whereas Figures~ \ref{fig:GIC_time}{d} and \ref{fig:GIC_time}{f} show the corresponding eastward GIC. It would be seen that both substations responded linearly to the input geoelectric field and responded correctly to three consecutive storms, both in the steep rise and the exponential decay phases. As in this algorithm, we did not consider the geographic inhomogeneity of the electric field yet; this step needs to account for site measurement of the geoelectric (or geomagnetic) field and reconstruct the electric field distribution at the ground, e.g., spherical elementary current system e.g,, SECS, Amm \cite{1997JGG}

\section{Application to Guangdong 500 kV power grid}

We here applied the \emph{Graph GIC} algorithm to the Guangdong 500 kV power grid in China. The current Guangdong power grid consists of a topology of 54 substations and 62 transmission lines, with no transformer information included. Therefore, the modeling of the Guangdong power grid considers substations as nodes and transmission lines as edges, and the impact of GIC on the substations is analyzed. The earthing resistance of the substations is shown in Table \ref{Tab:GD_substations}. Zhang et al. \cite{zhang2022} provided the geographic location (latitude and longitude) and the equivalent resistance of transmission lines of the Guangdong power grid.

\begin{figure}[ht]
  \centering

  \begin{minipage}[b]{0.45\textwidth}
    \centering
    \begin{picture}(0,0)
      \put(-100,25){\textbf{(a)}}
    \end{picture}
        \includegraphics[width=\textwidth]{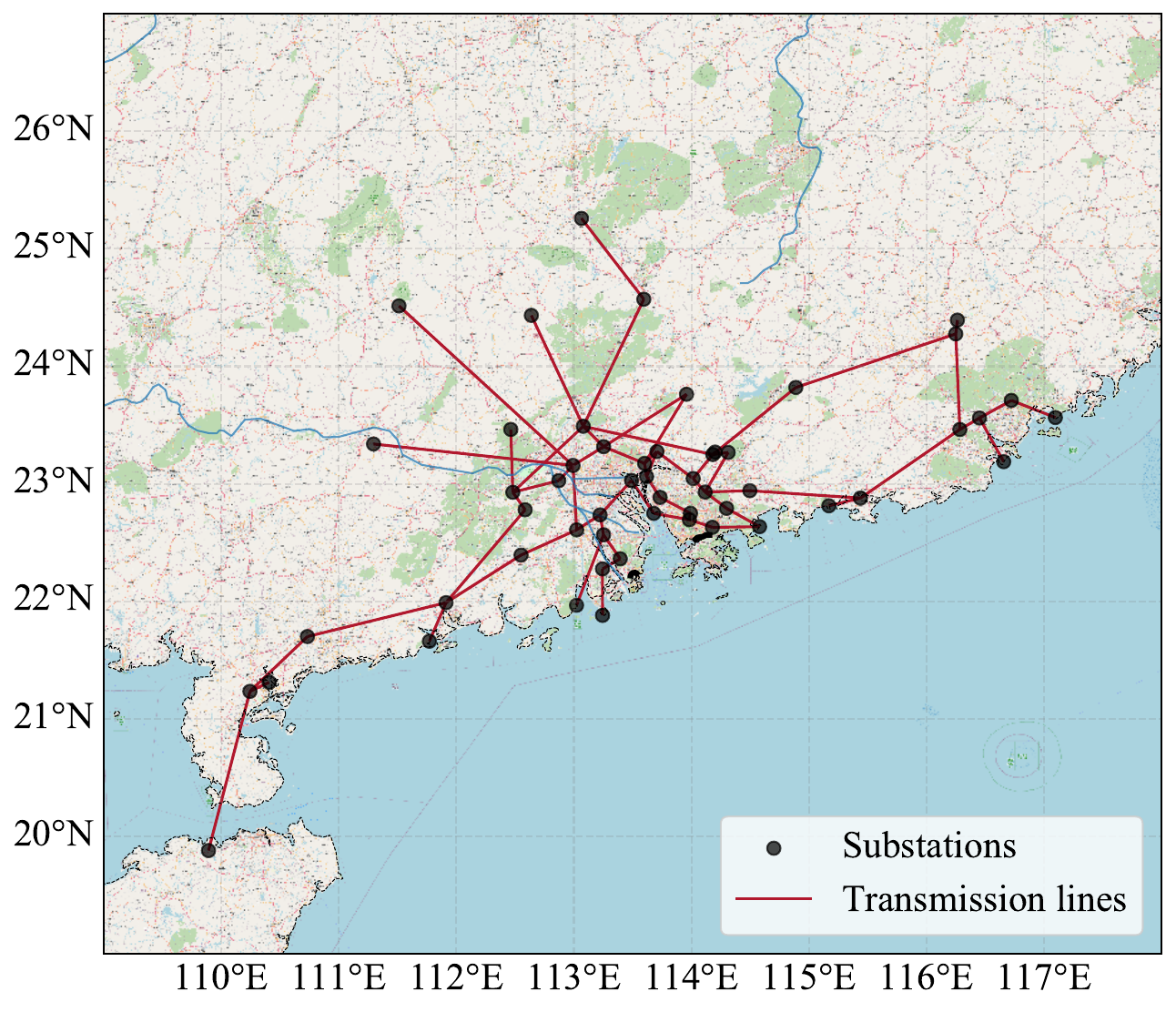}
  \end{minipage}
  \hfill
  \begin{minipage}[b]{0.45\textwidth}
    \centering
    \begin{picture}(0,0)
      \put(-100,0){\textbf{(b)}}
    \end{picture}
    \includegraphics[width=\textwidth]{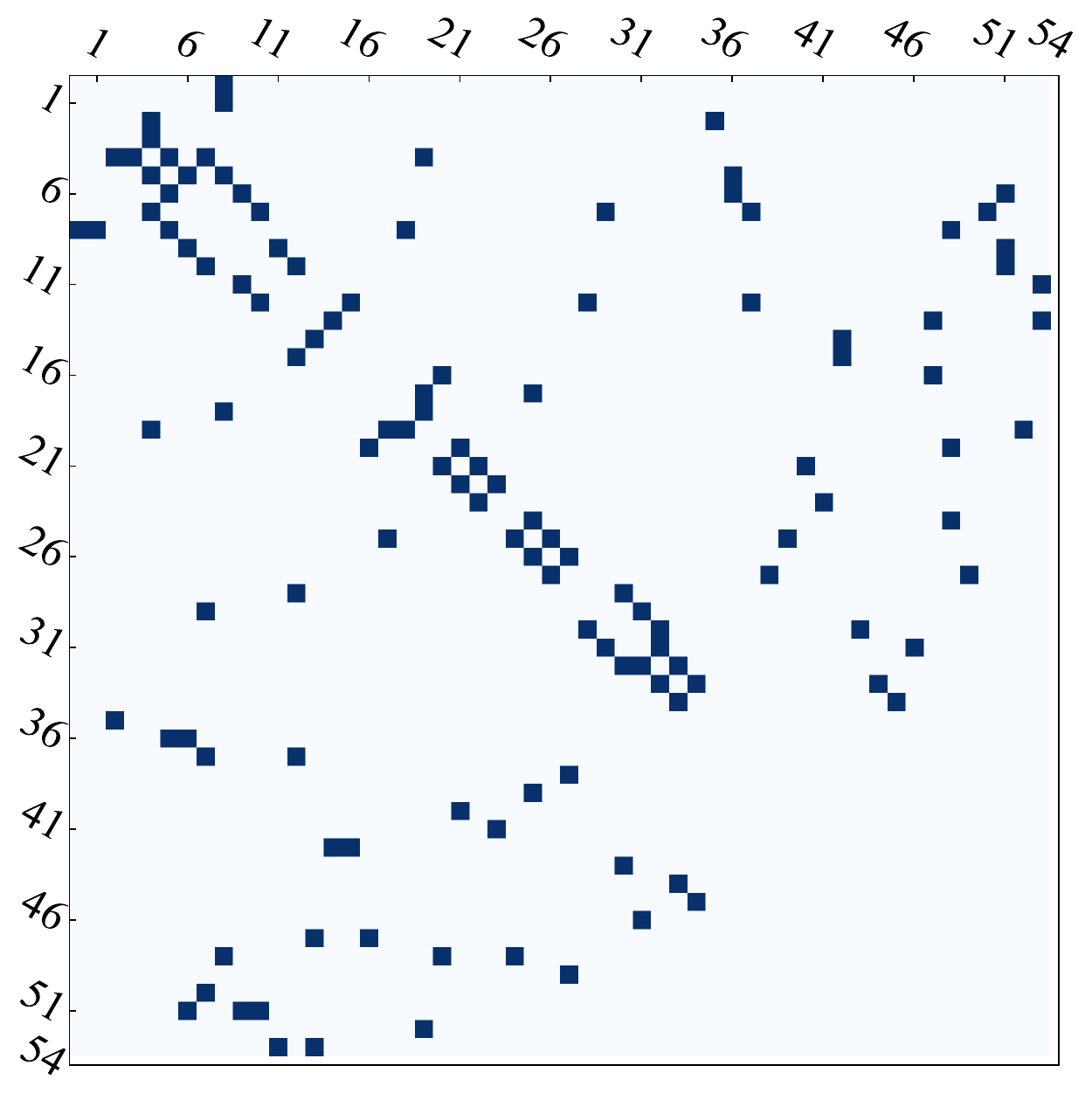}
  \end{minipage}
\caption{ Guangdong 500 kV power grid (a) and its adjacency matrix (b).}
    \label{fig:grid_guangdong}
\end{figure}

\begin{longtable}{cccccc}
\caption{Resistance information of 54 substations of Guangdong high-voltage power grid.} \\
\hline
No. & Name & Grounding resistances ($\Omega$) & No. & Name & Grounding resistances ($\Omega$) \\
\hline
\endfirsthead
\multicolumn{6}{c}%
{{\tablename\ \thetable{: Resistance information of 54 substations of Guangdong high-voltage power grid.}}} \\
\hline
No. & Name & Grounding resistances ($\Omega$) & No. & Name & Grounding resistances ($\Omega$) \\
\hline
\endhead
\hline \multicolumn{6}{r}{{Continued on next page}} \\ \hline
\endfoot

\endlastfoot
1 & Hezhou & 1.7 & 2 & Wuzhou & 1.7 \\
3 & Qujiang & 1.6 & 4 & Xianlingshan & 1.7 \\
5 & Huadu & 1.6 & 6 & Beijiao & 1.6 \\
7 & Zengcheng & 1.6 & 8 & Boluo & 1.6 \\
9 & Luodong & 1.4 & 10 & Shuixiang & 1.4 \\
11 & Hengli & 1.3 & 12 & Guancheng & 1.3 \\
13 & Dongguan & 1.5 & 14 & Pengcheng & 1.3 \\
15 & Shenzhen & 1.5 & 16 & Kunpeng & 1.6 \\
17 & Guangnan & 1.4 & 18 & Cangjiang & 1.7 \\
19 & Xijiang & 1.4 & 20 & Yandu & 1.6 \\
21 & Shunde & 1.4 & 22 & Xiangshan & 1.4 \\
23 & Guishan & 1.6 & 24 & Guoan & 1.6 \\
25 & Wuyi & 1.6 & 26 & Dieling & 1.6 \\
27 & Maoming & 1.6 & 28 & Gangcheng & 1.6 \\
29 & Huizhou & 1.5 & 30 & Shangzhai & 1.4 \\
31 & Maohu & 1.6 & 32 & Jiaying & 1.6 \\
33 & Rongjiang & 1.6 & 34 & Shantou & 1.6 \\
35 & Hanjiang & 1.6 & 36 & Pingshi & 1.6 \\
37 & Guangxu & 1.4 & 38 & Huixu & 1.4 \\
39 & Aoliyou & 1.6 & 40 & Yangxi & 1.6 \\
41 & Tonggu & 1.3 & 42 & Zhuhai & 1.6 \\
43 & Ling'ao & 1.3 & 44 & Honghaiwan & 1.6 \\
45 & Haimen & 1.5 & 46 & Tuolin & 1.6 \\
47 & Heshuyuan & 1.6 & 48 & shajiao & 1.2 \\
49 & Jiangmen & 1.6 & 50 & Fushan & 1.7 \\
51 & Echeng & 1.3 & 52 & Suidong & 1.0 \\
53 & Zhaoqing & 1.3 & 54 & Baoan & 1.3 \\

\bottomrule
\label{Tab:GD_substations}
\end{longtable}
Based on the location information of the substations and transmission lines, Figure~\ref{fig:grid_guangdong}{a} illustrates the Guangdong 500 kV power grid in a graph and overlaid on a map. Substations and transmission lines are denoted by nodes and edges, respectively. Figure~\ref{fig:grid_guangdong}{b} shows the adjacency matrix in a $54\times54$ format, revealing the connections between the substations. 

\subsection{Validation with the Uniform Electric Fields}
\label{sec:GD_uniform}

\begin{figure}[ht]
  \centering

    \begin{minipage}[b]{0.9\textwidth}
    \centering
    \begin{picture}(0,0)
      \put(-200,0){\textbf{(a)}}
    \end{picture}
        \includegraphics[width=\textwidth]{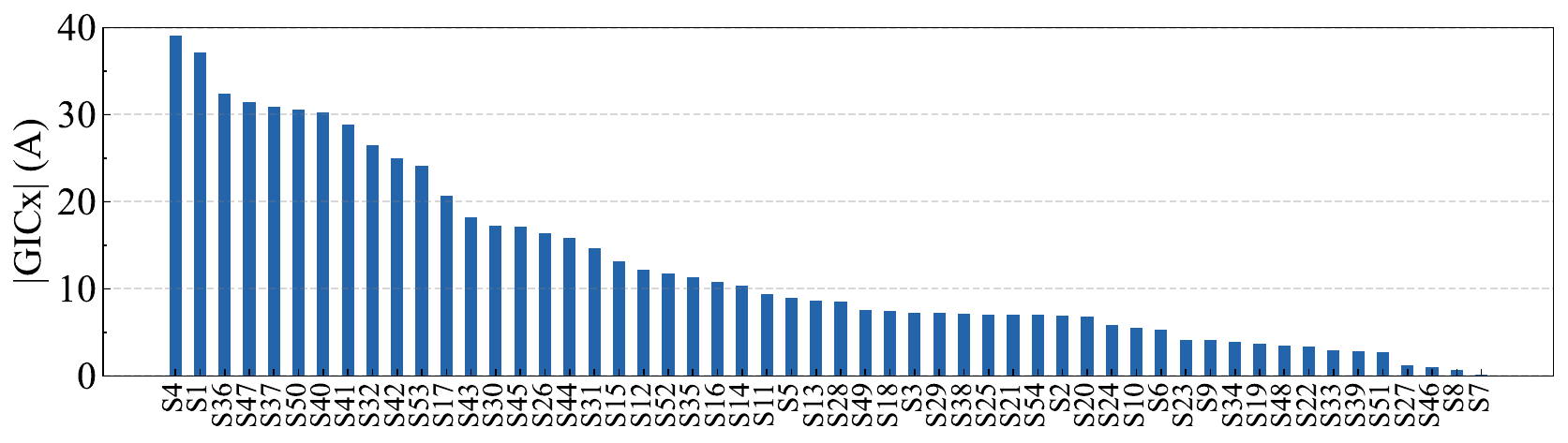}
    \end{minipage}  
\begin{minipage}[b]{0.9\textwidth}
        \centering
        \begin{picture}(0,0)
      \put(-200,0){\textbf{(b)}}
    \end{picture}
        \includegraphics[width=\textwidth]{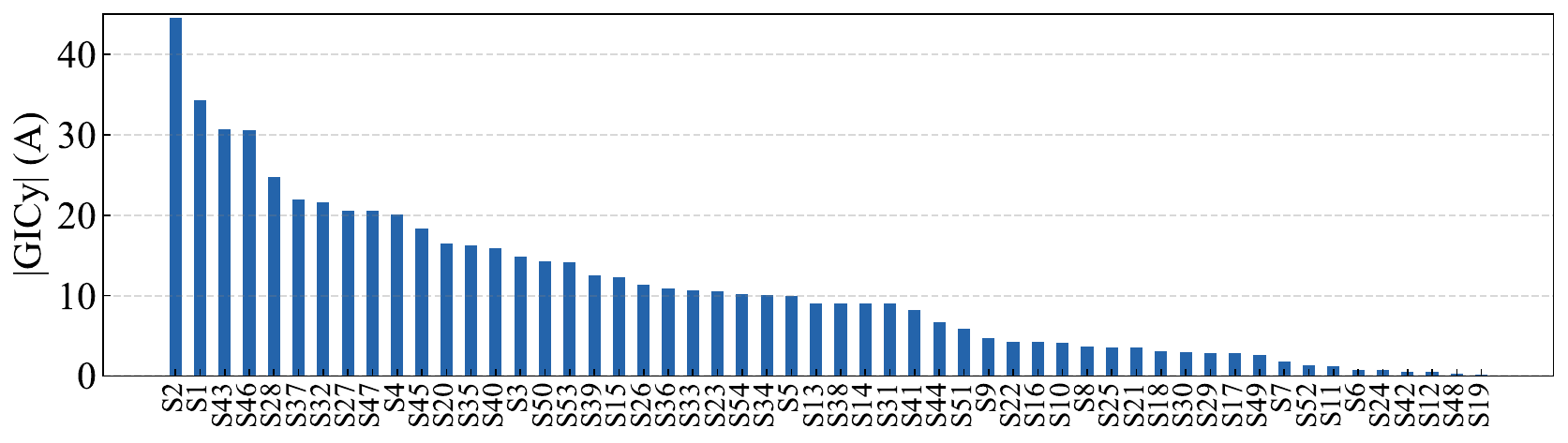}
\end{minipage}
\begin{minipage}[b]{0.9\textwidth}
    \centering
    \begin{picture}(0,0)
      \put(-200,0){\textbf{(c)}}
    \end{picture}
        \includegraphics[width=\textwidth]{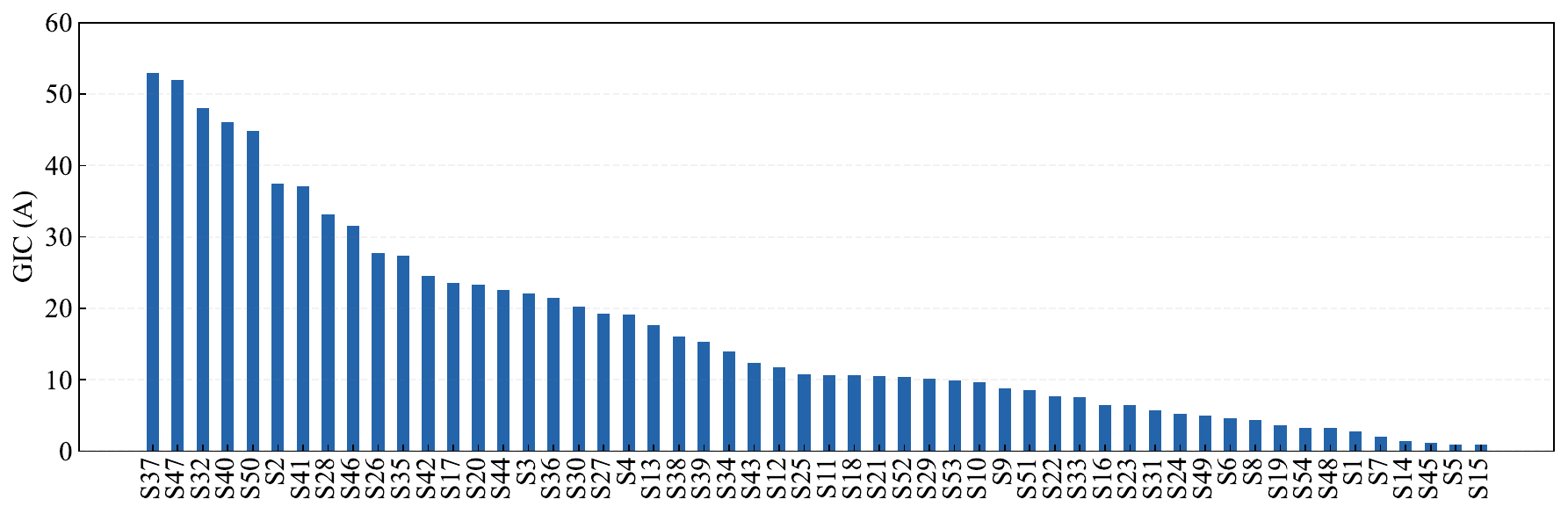}
    \end{minipage}  
\caption{(a) GIC amplitudes at 54 substations of the Guangdong 500 kV power grid under a northward geoelectric field. (b) GIC amplitudes under an eastward electric field. (c) \label{fig:GIC_GD} Total GIC amplitudes.}
\end{figure}

The GIC in the Guangdong 500 kV power grid are calculated under the assumption of a uniform 1 V/km electric field in both the northward and eastward directions. Based on the GIC levels at substations under a 1 V/km field, it becomes easier to estimate the GIC levels during geomagnetic storms \cite{viljanen1989}. Figure~\ref{fig:GIC_GD}(a) shows that the strongest northward GIC was obtained in Xianlingshan station (Substaion 4 or S4), with a current of 39.1 A flowing from the substation to the ground. additionally, there are five substations with GIC amplitudes greater than 30 A. Figure~\ref{fig:GIC_GD}(b) indicates the strongest eastward GIC was found at Wuzhou station, with a magnitude of 44.5 A flowing from the ground to substation. Four substations exhibit GIC amplitudes exceeding 30 A. Figure~\ref{fig:GIC_GD}(c) shows the GIC magnitudes at 54 substations under a 1 V/km electric field in the northward and eastward directions (only the magnitudes are shown).

According to Table S1 of Supporting Information, Hezhou station (S1) is connected to Luodong station (S9), which is $212.6\;$km in length. It forms a transmission line with an angle of $-47.5^\circ$ from the north, so it has a strong northward and eastward GIC. Similarly, the GIC at Xianlingshan station (S4) is twice as strong northward as eastward, mainly because the transmission line there is more aligned with the north-south axis. So, the alignment with the input geoelectric field and length of the transmission line for integration has a determinant effect on the GIC magnitude. 

Actually, At large "northward" and "eastward" currents, the resulting GIC in the transformer may become small. And the total GIC provides a more accurate representation of the combined GIC intensity flowing through the substations under both electric fields \cite{boteler1998}. As shown in Figure 8(c), both Hezhou station (S1) and Guangxu station (S37) exhibit large GIC under both northward and eastward electric fields. However, the total GIC at S1 is nearly zero, while s37 has the highest total GIC, with a magnitude of 44.5 A.

Under 1 V/km electric fields in both the northward and eastward directions, a significant amount of GIC, exceeding 50A, can flow through the power grid nodes. This indicates a substantial safety risk to the grid during geomagnetic storms.

\subsection{During Geomagnetic Storms}

\begin{figure}[ht]
\centering
\includegraphics[width=\textwidth]{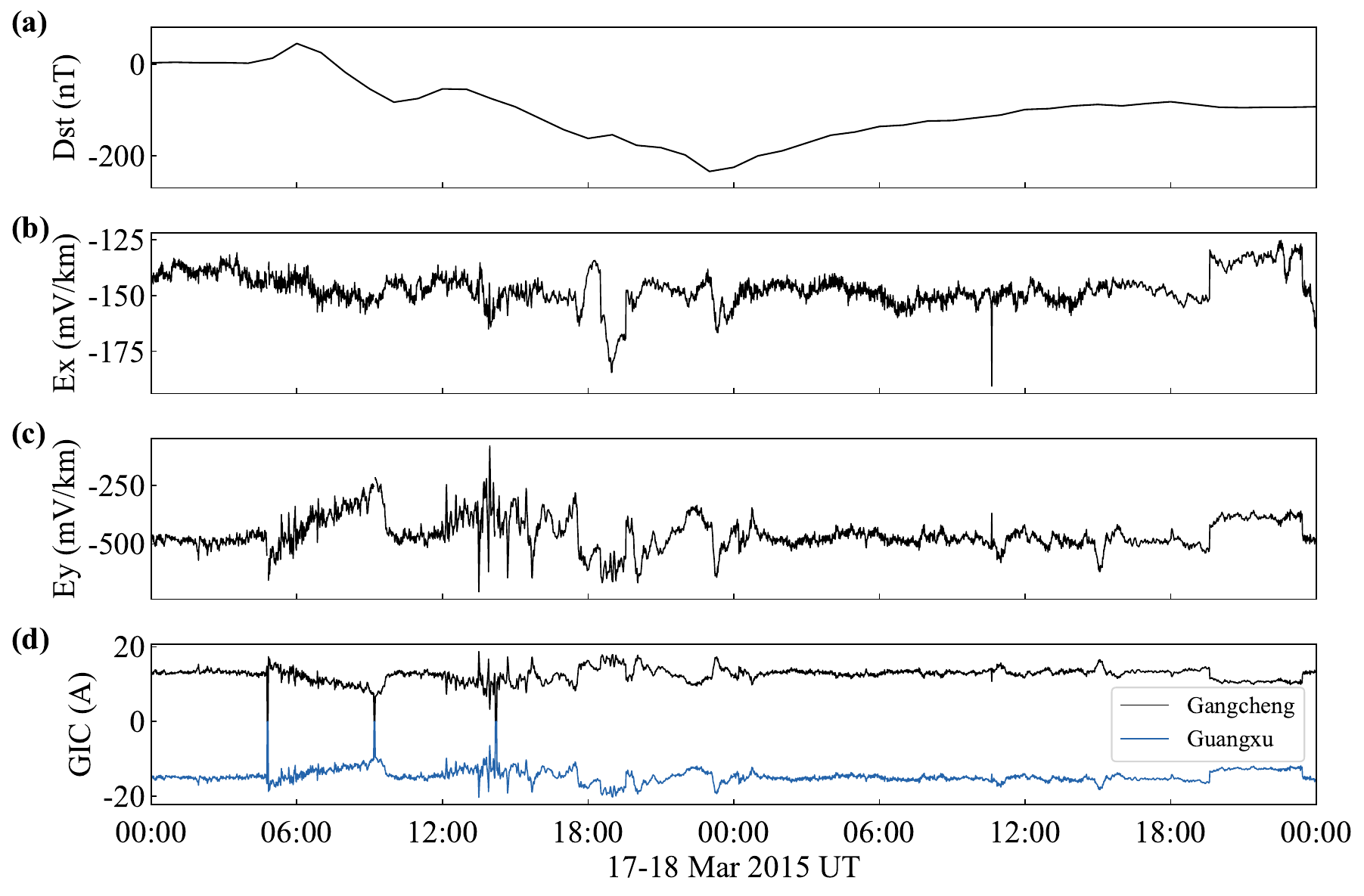}
\caption{(a) Dst from 00:00 UT 17 March to 24:00 UT 18 March 2015. (b)(c) Northward (Ex) and eastward (Ey) geoelectric fields measured at the Zhaoqing Station  (d) GIC calculated at the Gangcheng substation (S28) and Guangxu substation (S37). \label{fig:2015}}
\end{figure}

\begin{figure}[ht]
\centering
\includegraphics[width=\textwidth]{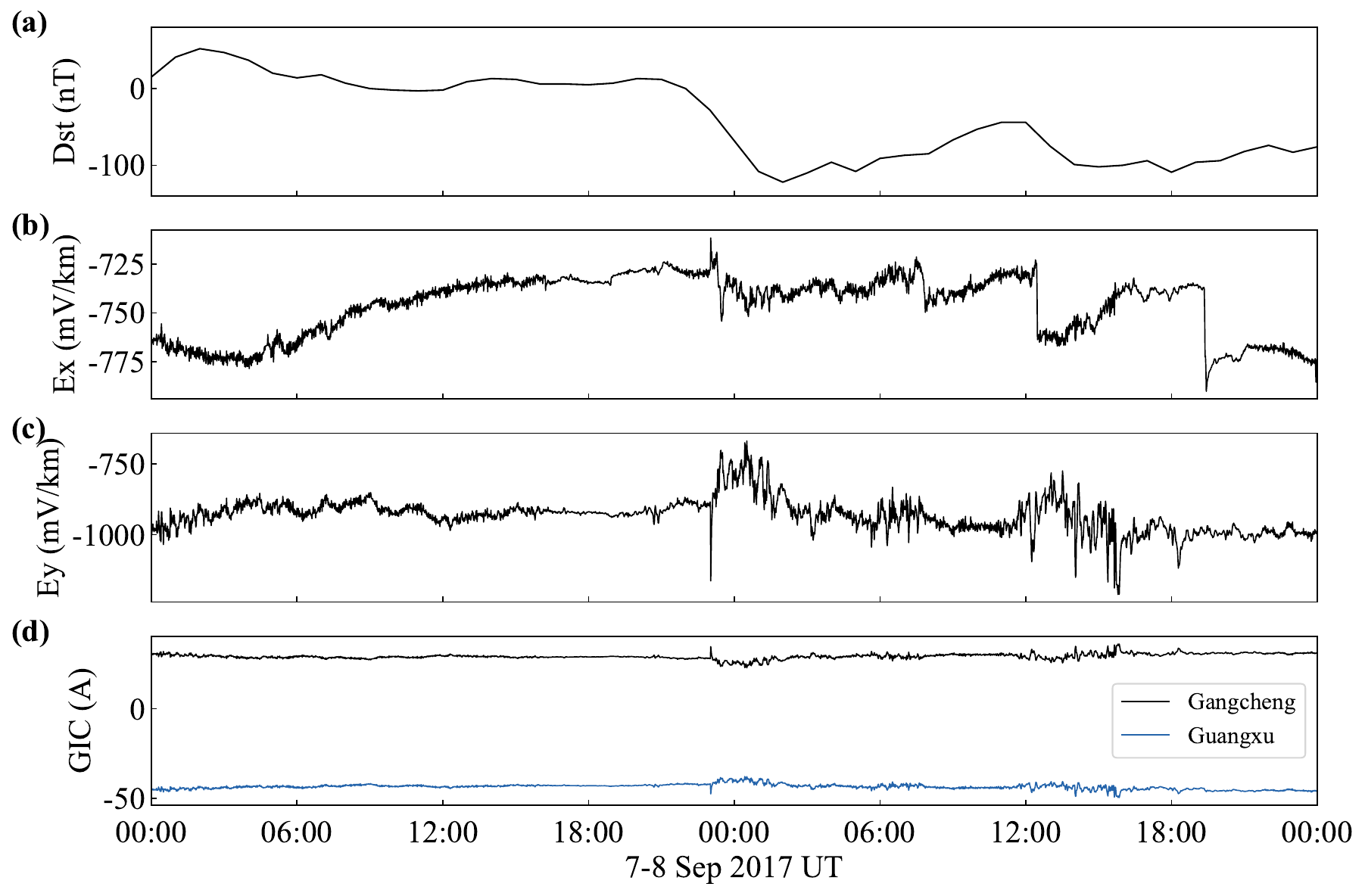}
\caption{(a) Dst from 00:00 UT on 7 September to 24:00 UT on 8 September 2017
(b)(c) Northward (Ex) and eastward (Ey) geoelectric field measured at the Zhaoqing Station. (d) GIC calculated at the Gangcheng substation (S28) and Guangxu substation (S37). \label{fig:2017}}
\end{figure}

To calculate the GIC under a geomagnetic storm, we used the geomagnetic field measured at the Zhaoqing station as the input to the power grid model. Zhaoqing station is a site for geoelectric vector field measurement as a part of the Chinese Meridian project {https://data.meridianproject.ac.cn/}. It provides both northward (Ex) and eastward components (Ey) of the geoelectric field. As this was the only geoelectric field measurement site, we assume that the Guangdong region was subject to a uniform but time-varying geoelectric field. In this study, we selected two geomagnetic storms: the 17 March 2015 storm with a peak Dst index of -234 nT and the 7 September 2017 storm with a peak Dst index of -122 nT. Figure~\ref{fig:2015}a and Figure~\ref{fig:2017}a show the variation of the Dst index for two magnetic storms, respectively.

Figure~\ref{fig:2015} depicts the time series of the geoelectric field monitored at the Zhaoqing geostation from March 17 to 18, 2015, along with the GIC at the Gangcheng station (S28) and Guangxu station (S37). For two substations, the GIC flowing into the substations reached their maximum at 13:31 on September 7, 2017, with values of 18.73 A and 20.29 A, respectively. Positive values indicate that the GIC flows from the ground into the transformer's neutral point. In contrast, negative values indicate that the GIC flows from the transformer neutral point into the ground.

Figure~\ref{fig:2017} shows the time series of the geoelectric field monitored at the Zhaoqing Geomagnetic Station from September 7 to 8, 2017, along with the GIC at the Gangcheng station (S28) and Guangxu station (S37). For two substations, the GIC flowing into the substations reached their maximum at 15:52 on March 18, 2015, with values of 36.23 A and 49.62 A, respectively.

\section{Conclusions and Discussion}

The geomagnetically induced current (GIC) is crucial in assessing the safety of substations and transformers in high-voltage power networks. This paper utilized advanced techniques, including network graph theory,  sparse matrices and parallel computing, to model GIC during geomagnetic storms in high-voltage power grids, referred to as the ``Graph-GIC" model. The model inputs geoelectric field data (which can be obtained through geomagnetic field calculations or geoelectric field monitoring stations) and outputs GIC at the substations and transformer windings. Furthermore, this paper computed the GIC magnitudes in the Guangdong power grid during both uniform electric field conditions and magnetic storm events. It analyzed the impact of the power grid structure on GIC.

Our modelling implemented an engineering model of grid GIC based on the LPm method \cite{pirjola2022}, integrating with network graph theory to depict the grid's topology. We depict the neutral nodes and buses as nodes and transmission lines as edges, characterizing this grid as an undirected graph and offering a clearer understanding of grid node connections. Additionally, the adjacency matrix of grid nodes and voltages of nodes were stored and computed in the form of a sparse matrix to enhance the efficiency of GIC computation. In addition, parallel computing is employed to calculate the time series of grid GIC. Increasing the number of cores and reducing the proportion of the serial part in the parallel computing process can significantly improve computational efficiency.

Our study conducted software testing using a benchmark model, which revealed that both substations and transformers exhibited extremely low errors, confirming the software's high accuracy. The test results indicate that the GIC error of 8 substations is less than 1\%. The maximum GIC error of transformer windings is 1.36\%, and other values are less than 1\%. Additionally, we simulated the time series of grid GIC, demonstrating a linear relationship between GIC magnitude and electric field intensity. 

The study computed and analyzed the GIC in the Guangdong power grid under uniform electric field conditions. The results indicate a positive correlation between line lengths and GIC magnitudes, and transmission lines aligned with the direction of the geoelectric field increase the GIC amplitude in the connected substation. Additionally, the GICs at Gangcheng station (S28) and Guangxu station (S37) rank among the highest despite the shorter line lengths, which aligns well with the corner effect. This study assessed the GIC magnitudes in the Guangdong power grid by integrating data from Zhaoqing geostation during two geomagnetic storm events on March 17, 2015 and September 6, 2017. The geoelectric field data from the two geomagnetic storm events indicate that the Ey (eastward) component is larger than the Ex (northward) component, suggesting that Ey reacts more sensitively to magnetic field disturbances. The analysis involved evaluating the GIC  magnitudes at substations, considering the intensity of the geomagnetic field and node responses. The time series analysis of GIC at Gangcheng station (S28) and Guangxu station (S37) indicates a linear correlation between GIC magnitude and geoelectric field, suggesting that higher geoelectric field values result in stronger GIC responses at the substations.

Furthermore, the extremely severe geomagnetic storm event on September 6, 2017, resulted in remarkably high GIC magnitudes in the Guangdong power grid. Despite the lower DST index during the 2017 event compared to the 2015 event, this discrepancy is primarily attributed to the proportionality between node GIC magnitudes and the first-order time derivative of the geomagnetic field. The main findings of this study hold significant theoretical and practical implications for GIC modelling and subsequent prevention techniques in low-latitude regions.

\section*{Acknowledgments}
W.C and D.Y were supported by the National Natural Science Foundation of China (NSFC,12173012, 12473050), the Guangdong Natural Science Funds for Distinguished Young Scholars (2023B1515020049), and the Shenzhen Key Laboratory Launching Project (No. ZDSYS20210702140800001). The authors acknowledge using geoelectric field data from the Chinese Meridian Project. S.P. acknowledges support from the projects C16/24/010 (C1 project Internal Funds KU Leuven), G0B5823N and G002523N (WEAVE) (FWO-Vlaanderen), 4000145223 (SIDC Data Exploitation (SIDEX2), ESA Prodex), and Belspo project B2/191/P1/SWiM.
\section*{Data Availability Statement}
The geoelectric field data used in this paper are available from the National Space Science Data Center:  \url{https://vsso.nssdc.ac.cn/nssdc_zh/html/task/ziwu.html}.

\bibliography{Graph_GIC_A_Smart_and_Parallelized_Geomagnetically_Induced_Current_Modelling_Algorithm_Based_on_Graph_Theory_for_Space_Weather_Applications}

\begin{thebibliography}{10}

\bibitem{albertson1981}
V.~Albertson, J.~Kappenman, N.~Mohan, and G.~Skarbakka.
\newblock Load-flow studies in the presence of geomagnetically-induced currents.
\newblock {\em IEEE transactions on power apparatus and systems}, PAS-100(2):594--607, 1981.

\bibitem{1997JGG}
O.~{Amm}.
\newblock {Ionospheric Elementary Current Systems in Spherical Coordinates and Their Application.}
\newblock {\em Journal of Geomagnetism and Geoelectricity}, 49(7):947--955, Jan. 1997.

\bibitem{boteler2014}
D.~Boteler and R.~Pirjola.
\newblock Comparison of methods for modelling geomagnetically induced currents.
\newblock In {\em Annales Geophysicae}, volume~32, pages 1177--1187. Copernicus Publications G{\"o}ttingen, Germany, 2014.

\bibitem{boteler1998}
D.~Boteler, R.~Pirjola, and H.~Nevanlinna.
\newblock The effects of geomagnetic disturbances on electrical systems at the earth's surface.
\newblock {\em Advances in Space Research}, 22(1):17--27, 1998.

\bibitem{boteler2019}
D.~H. Boteler.
\newblock A 21st century view of the march 1989 magnetic storm.
\newblock {\em Space Weather}, 17(10):1427--1441, 2019.

\bibitem{NetworkX-python}
A.~A. Hagberg, D.~A. Schult, and P.~J. Swart.
\newblock Exploring network structure, dynamics, and function using networkx.
\newblock In G.~Varoquaux, T.~Vaught, and J.~Millman, editors, {\em Proceedings of the 7th Python in Science Conference}, pages 11 -- 15, Pasadena, CA USA, 2008.

\bibitem{horton2012}
R.~Horton, D.~Boteler, T.~J. Overbye, R.~Pirjola, and R.~C. Dugan.
\newblock A test case for the calculation of geomagnetically induced currents.
\newblock {\em IEEE Transactions on power delivery}, 27(4):2368--2373, 2012.

\bibitem{kappenman1989}
J.~G. Kappenman.
\newblock Effects of geomagnetic disturbances on power systems [panel session-pes summer meeting, july 12, 1989 long beach, california].
\newblock {\em IEEE Power Engineering Review}, 9(10):15--20, 1989.

\bibitem{Kappernman1990}
J.~Kappernman and V.~Albertson.
\newblock Bracing for the geomagnetic storms.
\newblock {\em IEEE Spectrum}, 27(3):27--33, 1990.

\bibitem{lehtinen1985}
M.~Lehtinen.
\newblock Currents produced in earthed conductor networks by geomagnetically-induced electric fields.
\newblock {\em Ann. Geophys.}, 3(4):479--484, 1985.

\bibitem{liu2016}
L.~Liu, X.~Ge, W.~Zong, Y.~Zhou, and M.~Liu.
\newblock Analysis of the monitoring data of geomagnetic storm interference in the electrification system of a high-speed railway.
\newblock {\em Space Weather}, 14(10):754--763, 2016.

\bibitem{lundby1985}
S.~Lundby, B.~Chapel, D.~Boteler, T.~Watanabe, and R.~Horita.
\newblock Occurrence frequency of geomagnetically induced currents a case study on a bc hydro 500kv power line.
\newblock {\em Journal of geomagnetism and geoelectricity}, 37(12):1097--1114, 1985.

\bibitem{marti2012}
L.~Marti, A.~Rezaei-Zare, and A.~Narang.
\newblock Simulation of transformer hotspot heating due to geomagnetically induced currents.
\newblock {\em IEEE Transactions on Power Delivery}, 28(1):320--327, 2012.

\bibitem{molinski2002}
T.~S. Molinski.
\newblock Why utilities respect geomagnetically induced currents.
\newblock {\em Journal of atmospheric and solar-terrestrial physics}, 64(16):1765--1778, 2002.

\bibitem{ngwira2013}
C.~M. Ngwira, A.~Pulkkinen, F.~D. Wilder, and G.~Crowley.
\newblock Extended study of extreme geoelectric field event scenarios for geomagnetically induced current applications.
\newblock {\em Space Weather}, 11(3):121--131, 2013.

\bibitem{pilipenko2021}
V.~Pilipenko.
\newblock Space weather impact on ground-based technological systems.
\newblock {\em Solar-Terrestrial Physics}, 7(3):68--104, 2021.

\bibitem{pirjola2000}
R.~Pirjola.
\newblock Geomagnetically induced currents during magnetic storms.
\newblock {\em IEEE transactions on plasma science}, 28(6):1867--1873, 2000.

\bibitem{Pirjola2008}
R.~Pirjola.
\newblock Effects of interactions between stations on the calculation of geomagnetically induced currents in an electric power transmission system.
\newblock {\em Earth, Planets and Space}, 60(7):743--751, 2008.

\bibitem{pirjola2022}
R.~J. Pirjola, D.~H. Boteler, L.~Tuck, and S.~Marsal.
\newblock The lehtinen--pirjola method modified for efficient modelling of geomagnetically induced currents in multiple voltage levels of a power network.
\newblock {\em Annales Geophysicae}, 40(2):205--215, 2022.

\bibitem{press2007numerical}
W.~H. Press, S.~A. Teukolsky, W.~T. Vetterling, and B.~P. Flannery.
\newblock {\em Numerical Recipes 3rd Edition: The Art of Scientific Computing}.
\newblock Cambridge University Press, 3rd edition, 2007.

\bibitem{price2002}
P.~R. Price.
\newblock Geomagnetically induced current effects on transformers.
\newblock {\em IEEE transactions on power delivery}, 17(4):1002--1008, 2002.

\bibitem{pulkkinen2012}
A.~Pulkkinen, E.~Bernabeu, J.~Eichner, C.~Beggan, and A.~Thomson.
\newblock Generation of 100-year geomagnetically induced current scenarios.
\newblock {\em Space Weather}, 10(4), 2012.

\bibitem{STOTT198719}
B.~Stott and O.~Alsaç.
\newblock An overview of sparse matrix techniques for on-line network applications.
\newblock {\em IFAC Proceedings Volumes}, 20(6):19--25, 1987.
\newblock IFAC Symposium on Power Systems and Power Plant Control, Beijing, PRC, 12-15 August 1986.

\bibitem{viljanen1989}
A.~Viljanen and R.~Pirjola.
\newblock Statistics on geomagnetically-induced currents in the finnish 400kv power system based on recordings of geomagnetic variations.
\newblock {\em Journal of geomagnetism and geoelectricity}, 41(4):411--420, 1989.

\bibitem{zhang2022}
J.~Zhang, Y.~Yu, W.~Chen, C.~Wang, Y.~Liu, C.~Liu, and L.~Liu.
\newblock Simulation of geomagnetically induced currents in a low-latitude 500 kv power network during a solar superstorm.
\newblock {\em Space Weather}, 20(4):e2021SW003005, 2022.

\end{thebibliography}

\bibliographystyle{abbrv}

\end{document}